\documentclass[10pt,twoside]{article}
\usepackage{graphicx}
\usepackage{psfig}
\usepackage{amssymb}

\def\trait{\noalign{\smallskip\hrule\smallskip}}

\makeindex

\begin{document}

\title{Trans-sonic propeller stage}

\markboth{S.B.\,Popov and M.E.\,Prokhorov}{Trans-sonic propeller stage}

\author{S.B. Popov$^{1,*}$, M.E. Prokhorov$^{1,**}$
\\[5mm]
\it $^1$ SAI MSU, Moscow, Russia\\
\it e-mail: $^{*}$~polar@sai.msu.ru, $^{**}$~mike@sai.msu.ru \\
}

\date{}
\maketitle

\thispagestyle{empty}

\begin{abstract}
\noindent

We follow the approach used by Davies and Pringle (1981) %~\cite{DP81} %(1981) 
and discuss the trans-sonic substage of the propeller regime.
This substage is intermediate between the supersonic and subsonic propeller 
substages. In the trans-sonic regime an envelope around a magnetosphere 
of a neutron star  passes through a kind of a reorganization process.
The envelope in this regime
consists of two parts. In the bottom one turbulent motions are 
subsonic. Then at some distance $r_\mathrm{s}$ the turbulent velocity becomes
equal to the sound velocity. During this substage the boundary $r_\mathrm{s}$ 
propagates   outwards till it reaches the outer boundary, 
and so the subsonic regime starts.

We found that the trans-sonic substage is unstable, so the transition 
between supersonic and subsonic substages proceeds on the dynamical time scale.
For realistic parameters this time is in the range from weeks to years.

\end{abstract}

\section{Introduction} 
 
Observational appearances of a neutron star (NS)  
are mainly determined by its interaction  with the surrounding plasma. 
The following main stages (regimes) can be distingushed
(see a very detailed description in Lipunov 1992 %\cite{Lip92} 
or in Lipunov et al. 1996) %\cite{LPP96}):

\begin{itemize}
\item
\textit{Ejector}. At this stage plasma is swept away by 
low-frequency electromagnetic radiation or/and by a flow 
of relativistic particles. Matter is stopped further than 
the so-called light cylinder radius $r_{\ell}$.

\item
\textit{Propeller}. If a NS is in the propeller regime 
than matter can penetrate inside $r_{\ell}$, 
but it is stopped by a rapidly rotating magnetosphere of the NS.

\item
\textit{Accretion}. Finally the NS is slowed down and the 
centrifugal barrier disappeares, 
so if matter cools fast enough then it
can fall down onto the surface of the compact object.

\end{itemize}

Normally a NS is born at the stage of ejection 
(a radio pulsar is a classical example of a NS at the ejector stage). 
Then as the spin period increases the NS passes
propeller and accretor stages. For NSs with large ($>$400 km/s) 
spatial velocities another stage  --- \emph{Georotator} ---  can appear.

\begin{figure*}
\centerline{\hbox{
\psfig{figure=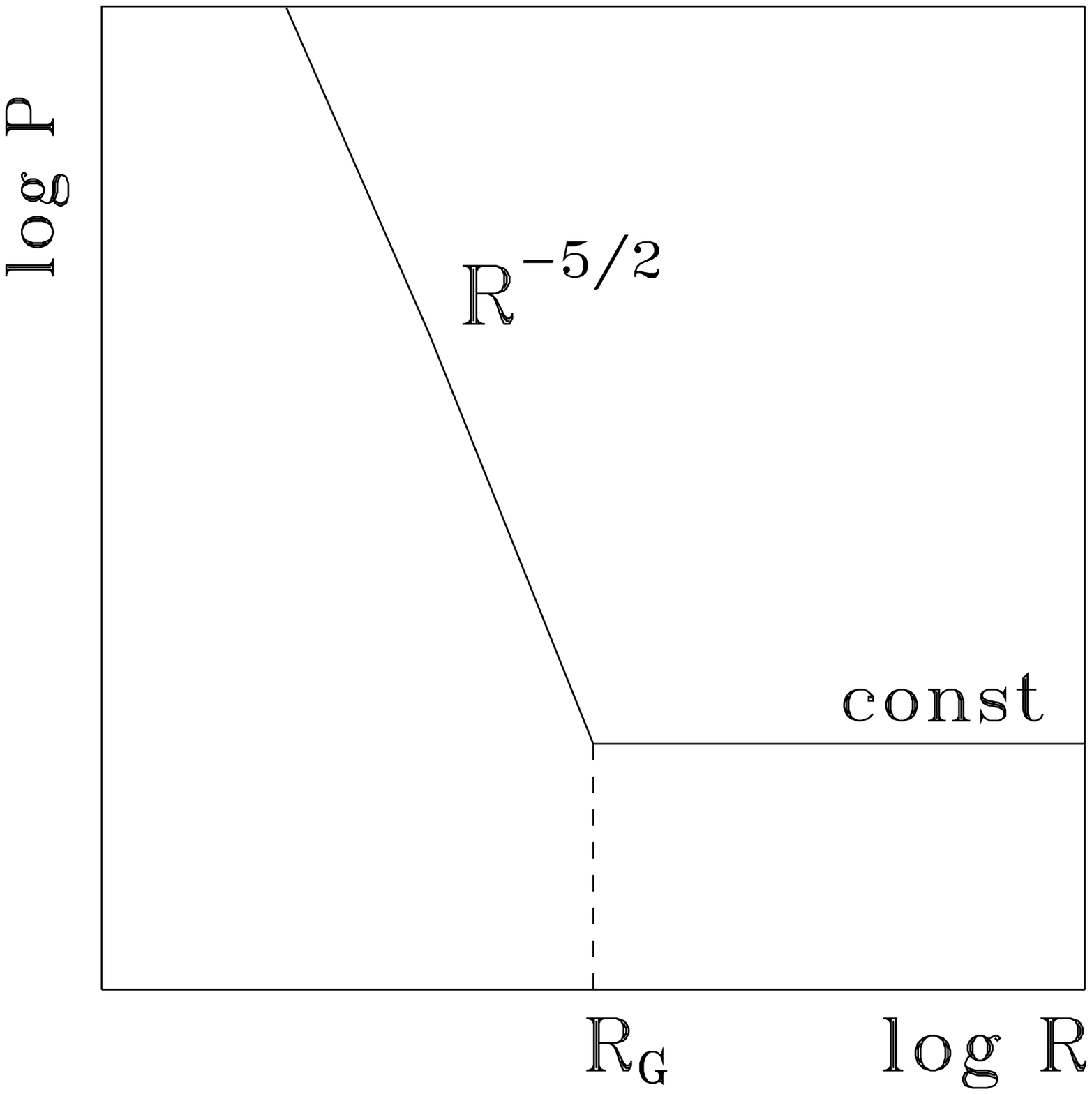,width=0.45\textwidth}
\hss
\psfig{figure=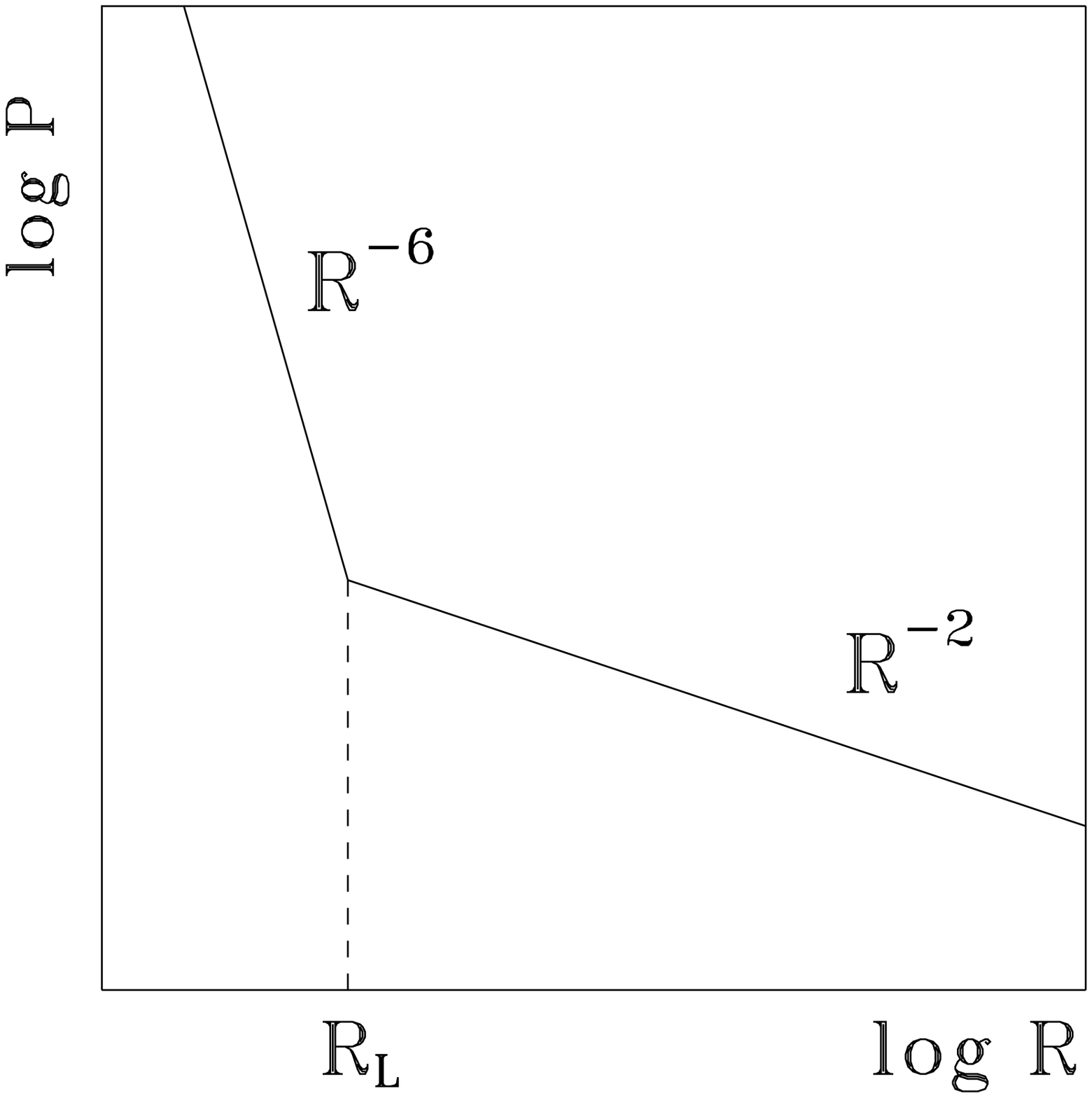,width=0.45\textwidth}
}}
\caption{Pressure vs. radius. Left panel: external pressure. 
Right panel: internal pressure.}
\label{fig1}
\end{figure*}

In a simplified model it is possible to define 
transitions between different stages by comparing 
external and internal pressure (Fig.~\ref{fig1}). 
The external one can be roughly approximated as the ram pressure of a 
flow of the interstellar medium (far from the NS) 
or as the pressure of matter falling down onto the NS in its 
gravitational field 
(for distances smaller than the so-called gravitational capture radius 
$r_\mathrm{G}$).
The internal one inside the light cylinder $r_{\ell}$ can be estimated 
as a pressure connected with the magnetic field of a NS. 

In this paper we discuss the propeller stage.
An existence of it was recognized long ago 
(initially by Shvartsman (1970) %\cite{sh70} 
and later on by Illarionov and Sunyaev(1975). %\cite{is75}).
However still  this stage is not well understood.
Here we introduce and discuss a new substage of this regime of 
magneto-rotational evolution of NSs.

\section{General features of the propeller stage}

%In \cite{DFP79,DP81} the authors
Davies et al. (1979) and Davies, Pringle (1981)
note that at the stage of propeller there can be a 
significant energy release at the magnetospheric boundary. 
Energy can be large enough to form a kind of a turbulent 
quasistatic atmosphere. These authors distinguished 
three substages of the propeller regime. 

\begin{enumerate}
\item
\textit{Very rapid rotator}:
\begin{equation}
c_\mathrm{s}(r_\mathrm{in}) \simeq r_\mathrm{in}\Omega \gg v_\mathrm{ff}\,.
\end{equation}
here  $c_\mathrm{s}$ -- sound velocity, $\Omega =2\pi/P$ -- spin frequency,
$v_\mathrm{ff}$ -- free-fall velocity.

\item
\textit{Supersonic propeller}:
\begin{equation}
r_\mathrm{in}\Omega  \gg c_\mathrm{s}(r_\mathrm{in})\,.
\end{equation}

\item
\textit{Subsonic propeller}:
\begin{equation}
r_\mathrm{in}\Omega  \ll c_\mathrm{s}(r_\mathrm{in})
\end{equation}
and
\begin{equation}
v_\mathrm{t}(r)< c_\mathrm{s}(r) \qquad\mathrm{in}\quad 
	r_\mathrm{in}<r< r_\mathrm{out}= r_\mathrm{G}\,,
\end{equation}

\end{enumerate}
here $r_\mathrm{in}$ and $r_\mathrm{out}$ --- 
are internal and external radii of an envelope, 
$v_\mathrm{t}$ --- turbulent velocity.

We will not discuss the substage of the very rapid propeller here.
Mainly we focus on super- and subsonic substages and 
on the transition between them.
The supersonic regime
can be considered a {\it classical} propeller where accretion is
impossible due to a centrifugal barrier.
At the subsonic stage the magnetospheric (Alfven) 
radius is smaller than the corotation radius 
$r_\mathrm{c}$. Accretion does not start because temperature is too high
(see original discussion in Davies, Pringle, 1981 %\cite{DP81} 
and later proposals in Ikhsanov 2003). % \cite{Iks2003}).

Braking laws (for the spin-down) are different at 
different substages. In the supersonic regime 
energy loss rate does not depend on the spin period (Davies, Pringle 1981),
%(\cite{DP81}, 
however different formulas for the spin-down at this stage were suggested, see a list
for example in (Lipunov, Popov 1995). % \cite{lp95}). 
In the subsonic one the spin-down is always slower: $P \propto t$. 

\section{Why does the intermediate regime exist?}

In general 
supersonic and subsonic regimes  cover 
all possible values of the rotation velocity $\Omega=2\pi/P$. 
The supersonic propeller formally operates till  
$ r^{super}_\mathrm{in}\Omega  \geq c_\mathrm{s}(r^{super}_\mathrm{in})\,.$
Here $r^{super}_\mathrm{in}=r_\mathrm{G}^{2/9}r_\mathrm{M}^{7/9}$~
(Davies, Pringle 1981), %\cite{DP81},
$r_\mathrm{M}=(\mu^2/(\dot M \sqrt{2GM}))^{2/7}$ -- 
magnetospheric (Alfven) radius.

The subsonic regime is  on  when 

\begin{equation}
	r_\mathrm{in}^{sub}\Omega  \leq c_\mathrm{s}(r^{sub}_\mathrm{in})
\, \, {\rm and}\, \,
	 v_\mathrm{t}(r) \leq c_\mathrm{s}(r)
	\qquad\mathrm{for}\quad
	r^{sub}_\mathrm{in}\leq r\leq r_\mathrm{out}= r_\mathrm{G}\,,
\end{equation}
here $r^{sub}_\mathrm{in}=r_\mathrm{M}$ (Davies, Pringle 1981). %\cite{DP81}.
As for the subsonic stage
	$ v_\mathrm{t}(r^{sub}_\mathrm{in}) \simeq r^{sub}_\mathrm{in}\Omega
\,$
~and~
$v_\mathrm{t}/c_\mathrm{s} \propto r^{1/3}\,,$
then this regime is valid for
	$r_\mathrm{in}^{sub}\Omega  \leq c_\mathrm{s}(r^{sub}_\mathrm{in}) 
         \left({r^{sub}_\mathrm{in}}/{r_\mathrm{G}}\right)^{1/3}\,.$

Note the following properties of the stages.

\begin{enumerate}
\item
At both stages internal radii of an envelope $r_\mathrm{in}$ do not depend
on $\Omega$, and always $r^{super}_\mathrm{in} > r^{sub}_\mathrm{in}$.

\item
The structure of the envelope on the two substages is different.

\item
It is easy to check that the end of the supersonic substage and the beginning
of the subsonic one both correspond to:
\begin{equation}
\Omega=\sqrt{2GM} r_\mathrm{M}^{-7/6}r_\mathrm{G}^{-1/3},
\end{equation}

$$
P \sim 1.15 \, 10^4 {\rm s}\, \mu_{30}^{2/3}v_6^{1/3}\rho_{-24}^{-1/3}.
$$

\end{enumerate}

Even in the frame of Davies and Pringle model
an intermediate regime during which the structure of the envelope is 
reorganized is inevitable.
We call this intermediate stage  \textit{trans-sonic propeller}.
As we show below this is a short nonequilibrium
episode in a life of a NS, and the spin frequency
does not change significantly during this reorganization process.
%%END

\section{Trans-sonic propeller}

Let us consider the
structure of a quasistationary envelope at the intermediate trans-sonic 
propeller substage.
 In the atmosphere around such a NS 
$c_\mathrm{s}(r)\simeq v_\mathrm{ff}$~(Davies, Pringle 1981). %\cite{DP81}.
(This condition is valid also for super-- and subsonic propeller.)
Processes in the lower part of the atmosphere 
are similar to the ones on the subsonic stage:
\begin{equation}
	v_\mathrm{t}(r_\mathrm{in}) \simeq r_\mathrm{in}\Omega\,<c_\mathrm{s}.
\end{equation}

If the envelope (atmosphere) is adiabatic 
then for this bottom part of it the politropic index is equal to $n=3/2$ and 
\begin{equation}
	\rho(r) \propto r^{-3/2}\,,\qquad p(r) \propto r^{-5/2}\,.
\end{equation}
We assume following Davies and Pringle~(1981) %\cite{DP81}  
that the rotational energy of the NS is dissipated 
at the magnetospheric boundary and 
that it is transported outwards by turbulence. For such assumptions we have:
\begin{equation}
	v_\mathrm{t}(r) \propto r^{-1/6}
\end{equation}
and the turbulent Mach number is:
\begin{equation}
	\mathcal{M}_\mathrm{t}(r) \equiv 
         \frac{v_\mathrm{t}(r)}{c_\mathrm{s}(r)} 
         	\propto r^{1/3}\,.
\end{equation}

Till $\mathcal{M}_\mathrm{t}<1$, 
i.e. while $r<r_\mathrm{s}$ where $r_\mathrm{s}$ is the boundary between two
parts of the envelope
($r_\mathrm{in}<r_\mathrm{s}< r_\mathrm{G}$) 
the structure of the envelope is not changed. 
For large radii turbulence becomes supersonic. 
Small-scale shock waves are formed 
and they quickly dissipate part of the energy, 
so that turbulent velocity decreases down to the sound velocity. In the range
$r_\mathrm{s}<r<r_\mathrm{G}$
the envelope structure is different from the bottom part:
\begin{equation}
	\mathcal{M}_\mathrm{t}(r) \simeq 1\,,
\end{equation}
\begin{equation}
	\rho(r) \propto r^{-1/2}\,,\qquad p(r) \propto r^{-3/2}\,.
\end{equation}

In the outer part of the envelope 
physical conditions are similar to the ones on the supersonic substage.

To determine parameters of the whole atmosphere 
it is necessary to calculate the position of the boundary 
between two parts of the envelope, $r_\mathrm{s}$, 
and position of the inner boundary of the bottom part, $r_\mathrm{in}$
(during the transition it decreases from $r_\mathrm{G}^{2/9}r_\mathrm{M}^{7/9}$
to $r_\mathrm{M}$). 
To do it it is necessary to solve the following system of equations:

\begin{equation}
\left\{
\begin{array}{rl}
\displaystyle{ \frac{\mu^2}{8\pi}\frac{1}{r_\mathrm{in}^6} } & 
\displaystyle{= \frac{1}{2}\frac{\dot Mv_\infty}{4\pi r_\mathrm{G}^2}
  	\left(\frac{r_\mathrm{G}}{r_\mathrm{s}}\right)^{3/2}
  	\left(\frac{r_\mathrm{s}}{r_\mathrm{in}}\right)^{5/2} }
\\
\displaystyle{ \Omega\, r_\mathrm{in}^{7/6} r_\mathrm{s}^{-1/6} } & 
\displaystyle{= \sqrt{\frac{2GM\,}{r_\mathrm{s}}} }
\end{array}
\right.
\label{eq:system}
\end{equation}
However, the system is degenerate and each equation 
can be reduced to:
$r_\mathrm{s} \propto r_\mathrm{in}^{-7/2}$. If the following equation is
fulfilled:
\begin{equation}
\frac{\mu^2}{8\pi}= 
\left(\frac{1}{2}\frac{\dot Mv_\infty}{4\pi r_\mathrm{G}^{1/2}}\right) 
\frac{(2GM)^{3/2}}{\Omega^3}
\label{eq:sovm}
\end{equation}
then the system is compatible, i.e. for all $r_{in}$ in
the range $r_\mathrm{M} < r_\mathrm{in} <
	r_\mathrm{M}^{7/9}r_\mathrm{G}^{2/9} $ 
        There is some $r_\mathrm{s}$ that is 
a solution of the eq. ~\ref{eq:system}.

The compatibility condition is fulfilled at the end of the supersonic substage.
Later (during the transition) at any given moment (for any $\Omega$) 
left-hand side of  eq.\ref{eq:sovm} is smaller than the right-hand one. 
It means that the magnetospheric pressure
and the envelope pressure have the same dependences on $r_\mathrm{in}$,
but the latter one is always larger (the first equation of the system
\ref{eq:system}).

During the trans-sonic stage the period is not changing significantly
(a typical value is determined by eq.~6), 
so in terms of the rotational evolution the subsonic substage 
nearly immediatelly follows the supersonic one.
The spin-down law for the trans-sonic propeller 
is the same as for the subsonic regime.

An energy release during the transition stage is negligible.
Estimates for realistic isolated NS parameters give a value 
$\Delta E\lesssim 10^{30}$~erg.

\section{Discussion}

We want to note that calculation similar to the ones presented above
are just rough estimates. There are several reasons for that.

The first is connected with uncertainties in many parameters, 
even in their determinations. For example the accretion rate 
is usually taken as $\dot M= \pi r_\mathrm{G}^2 \rho_{\infty} v_{\infty}$.
This is just an estimate, and for different velocitites it can be different
from the actual value by a factor a few. 
Small changes is some parameters can lead to significant changes 
in others. For example Ikhsanov (2001)
%in the paper \cite{Iks2001} the author 
discusses the value of the critical period which determines the end
of the subsonic propeller stage (and so the accretion stage begins).  
The obtained value is different from the
one found in the original paper by Davies and Pringle (1981) %\cite{DP81} 
by a factor of 7.5.
Correspondently all time scales are also significantly changed.
But note, that this fact is due to a change in the magnetospheric radius 
($r_\mathrm{M}$) only by a factor of $\sim2$! 

The second one is connected with idealizations. Even if all parameters can be 
well defined, then it is necessary to take into account such details as
non-spherical form of the magnitosphere, inclination of the magnetic axis
relative to the spin axis,
angular moment in the infalling matter
(even for cases when the condition for disc formation is not fulfilled), etc.
For example even if $r_\mathrm{M}>r_\mathrm{c}$ part of the magnetosphere
is inside $r_\mathrm{c}$ as  the corotation radius is the
radius of a cylinder, not a sphere. In that sense the process of alignment
(see for example Regimbau, de Freitas Pacheco 2001 and Beskin et al. 1993)
%\cite{f2001} and \cite{b1993})
during the magneto-rotational evolution of a NS can be important in a
destiny of a NS.

A low rate 
accretion can proceed even at the propeller stage due to several reasons.
One of them is diffusion of plasma. Such form of accretion was discussed in 
details by Ikhsanov (2003).% \cite{Iks2003}. 
For long spin periods luminosity due
to such an accretion can be larger than the dissipation of the the rotational
energy on the boundary of the magnetosphere.

An important question is connected with the whole time of evolution prior
the accretor stage $t_\mathrm{A}$. Obviously

\begin{equation}
t_\mathrm{A}=t_\mathrm{E}+t_\mathrm{P},
\end{equation}
where $t_\mathrm{E}$ is the time which a NS spends at the stage of ejector,
and $t_\mathrm{P}$ is the duration of the propeller stage.
Even $t_\mathrm{E}$  is not well determined. Usually authors assume that
spin-down at this stage is determined by the magneto-dipole formula with the
braking index equal to 3. However, direct measurement for many radio pulsars
show that the braking index is smaller than 3. 
Also an evolution of the angle between spin and magnetic axes
is usually not taken into account.

As the propeller stage consists of the three substages then
it is necessary to write: 

\begin{equation}
t_\mathrm{P}=t_\mathrm{super}+
t_\mathrm{trans}+t_\mathrm{sub}.
\end{equation}

\begin{figure}[t]
\centerline{\psfig{figure=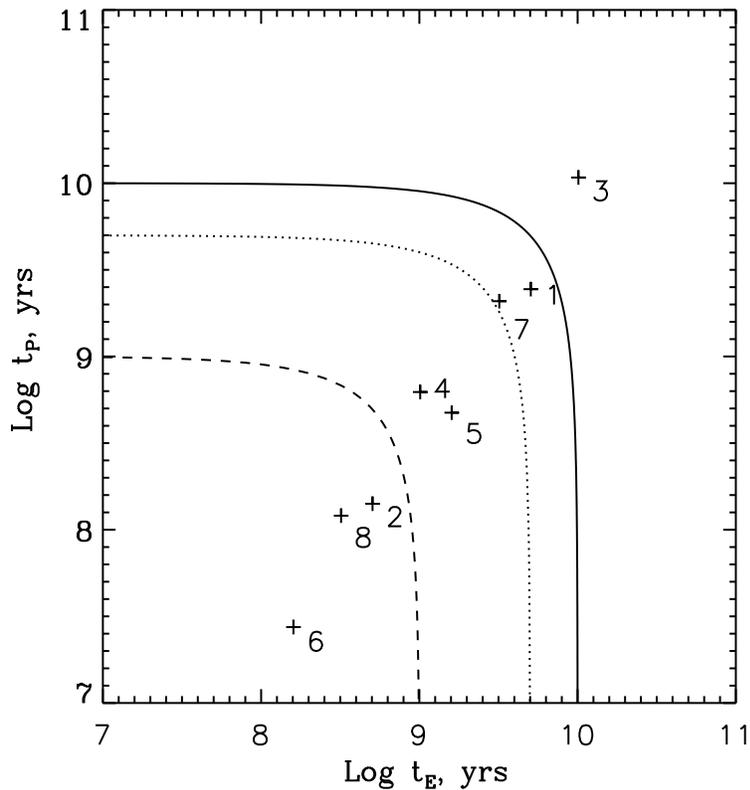,width=0.9\textwidth}}
\caption{$t_\mathrm{E}-t_\mathrm{P}$-diagram for isolated NSs.
Quasi-linear nature of the distribution of model points on the
log $t_\mathrm{E}$ -- log $t_\mathrm{P}$ plane was discussed in 
detail in (Popov 2004).}
% \cite{p2004}.}
\label{tetp}
\end{figure}

If $t_\mathrm{E}$ and $t_\mathrm{P}$ are determined
the fate of NSs for different parameters can be easily shown on
$t_\mathrm{E}-t_\mathrm{P}$-diagram suggested in (Popov 2004). %\cite{p2004}.
In the Fig.\ref{tetp} we show an example of such a plot.
For this illustration we assume that the accretion regime 
starts at the period:
\begin{equation}
P_\mathrm{br}=4.5\cdot10^7\mu_{30}^{16/21}\dot M_8^{-5/7} m^{-4/21},
\end{equation}
$\dot M_8=\dot M/10^8\, {\rm g s}^{-1}$, $m=M_\mathrm{NS}/M_{\odot}$.
Here we renormalize the value by Ikhsanov (2001).%\cite{Iks2001}.

Time scales are determined by (see Popov 2004):% \cite{p2004}):

\begin{equation}
t_\mathrm{E}=0.8\cdot 10^9 \mu_{30}^{-1}n^{-1/2}v_6 \, {\rm yr},
\end{equation}
here $n=\rho m_\mathrm{p}^{-1}$ -- interstellar medium number density, 
$m_\mathrm{p}$ -- proton mass.

\begin{equation}
t_\mathrm{super}=1.3\cdot 10^6 \mu_{30}^{-8/7}n^{-3/7}v_6^{9/7}\, {\rm yr},
\end{equation}
this is a very efficient spin-down suggested by Shakura (1975). %\cite{shak75},
so our estimate of $t_\mathrm{super}$ is in some sense a low bound.

\begin{equation}
t_\mathrm{trans}+t_\mathrm{sub}=10^3 \, \mu_{30}^{-2} m\, 
P_\mathrm{br}\, {\rm yr}.
\end{equation}
The mass of a NS is assumed to be $M_\mathrm{NS}=1.4\,M_{\odot}$.

In the plot we show lines for $t_\mathrm{E}+t_\mathrm{P}$ equal to 1, 5 and 
10 Gyrs. Eight symbols corresponds to eight combitations of $n, v, \mu$
(see the table).

\begin{table}
\caption{
$t_\mathrm{E}$ and $t_\mathrm{P}$ for typical values of 
$n$, $v$ and $\mu$
}
\begin{center}
\begin{tabular}{lccccc}
\trait
\multicolumn{1}{c}{Number} & $n$ & $v$, km s$^{-1}$ & $\mu_{30}$ &
log $t_\mathrm{E}$, yrs& log $t_\mathrm{P}$, yrs\\
\trait
1 & 0.1 & 20 & 1  &9.70 &9.39\\
2 & 0.1 & 20 & 10 & 8.70&8.15\\ 
3 & 0.1 & 40 & 1  & 10.0&10.0\\
4 & 0.1 & 40 & 10 & 9.01&8.79\\
5 & 1.0 & 20 & 1  & 9.20&8.68\\ 
6 & 1.0 & 20 & 10 &8.20 &7.44\\
7 & 1.0 & 40 & 1  &9.51 &9.32\\
8 & 1.0 & 40 & 10 & 8.51&8.08\\
\trait
\end{tabular}
\end{center}
\end{table}

\section{Conclusions}

In this paper we showed that an intermediate substage of the propeller regime
-- {\it the trans-sonic propeller} -- should exist. 
However, the stage is 
non-stationary and very short.
As it was shown above the existence of this intermediate stage 
does not change the timescale of the evolution prior to the accretor stage 
(see~Ikhsanov 2003 %\cite{Iks2003} 
for a discussion of the timescales). 
Other conclusions can be summarized as follows:

\begin{itemize}
\item
The intermediate trans-sonic propeller substage in unstable.

\item
The duration of the transition can be roughly estimated as
$\sim$$r_\mathrm{G}/v_\mathit{ff}$ 
(from weeks to years for realistic isolated NSs).

\item
The spin frequency is nearly unchanged during this transition.

\item
The energy release during the transition is small.

\end{itemize}

%--------------------------------------------------------

\bigskip

The work was partially supported by grants RBRF 04-02-16720 and 03-02-16068.

S.P. acknowleges a postdoctoral fellowship from the University of Padova
where most of this work has been carried out.

\end{document}